\documentclass[10pt,aps,prc,floatfix,twocolumn,nofootinbib,superscriptaddress,longbibliography]{revtex4-1}
\usepackage{graphicx,amsmath,amssymb,bm}
\usepackage{amsfonts}
\usepackage[dvipsnames]{xcolor}

\usepackage{soulpos}  

\usepackage[utf8]{inputenc}
\usepackage{verbatim}
\usepackage{float}
\usepackage[labelfont={small},font=small,subrefformat=parens,caption=false]{subfig}
\usepackage{cancel}
\usepackage{multirow}
\usepackage{array}
\usepackage{xparse}
\usepackage{xspace}

\usepackage{bigstrut}
\usepackage{physics}
\usepackage{color}
\usepackage{dsfont}
\usepackage{algorithm2e, algorithmic}

\usepackage{dcolumn}

\usepackage{xurl}
\usepackage[pdfencoding=auto, pdfpagelabels]{hyperref}
\definecolor{linkcolor}{rgb}{0,0,0.40} 
\hypersetup{%
    pdfsubject=Paper,
    pdfkeywords={nuclear astrophysics} {TOV equations} {reduced order models} {reduced basis methods},
    unicode = true,
    breaklinks = true,
    colorlinks = true,
    linkcolor = linkcolor,
    citecolor = linkcolor,
    menucolor = linkcolor,
    urlcolor = linkcolor
}

\usepackage{cellspace}
\usepackage[noabbrev]{cleveref}
\graphicspath{{./figures_draft/}}

\newcommand{\p}{\partial}
\definecolor{Joshpurple}{RGB}{155, 48, 255}

\newcommand{\orcid}[1]{\href{https://orcid.org/#1}{\includegraphics[scale=0.055]{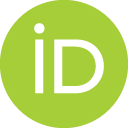}}}
\RestyleAlgo{ruled}

\definecolor{lightblue}{rgb}{.87,.95,.99}

\begin{document}

\preprint{APS/123-QED}

\title{Star Log-extended eMulation: a method for efficient computation of the Tolman-Oppenheimer-Volkoff equations}

\author{Sudhanva Lalit
\orcid{0000-0001-7758-492X}}
\email{lalit@frib.msu.edu}
\affiliation{%
 Facility for Rare Isotope Beams, Michigan State University, East Lansing, MI 48824, USA}%

\author{Alexandra C. Semposki
\orcid{0000-0003-2354-1523}}
\email{as727414@ohio.edu}
\affiliation{Department of Physics and Astronomy and Institute of Nuclear and Particle Physics, Ohio University, Athens, OH 45701, USA}

\author{Joshua M. Maldonado
\orcid{0009-0009-8872-2385}}
\email{jm998521@ohio.edu}
\affiliation{Department of Physics and Astronomy and Institute of Nuclear and Particle Physics, Ohio University, Athens, OH 45701, USA}

\date{\today}

\begin{abstract}
We emulate the Tolman-Oppenheimer-Volkoff (TOV) equations, including tidal deformability, for neutron stars using a new method based upon the Dynamic Mode Decomposition (DMD). This method, which we call Star Log-extended eMulation (SLM), utilizes the underlying logarithmic behavior of the differential equations to enable accurate emulation of the nonlinear system. We show predictions for well-known equations of state (EOSs) with fixed parameters using the SLM, accurately recreating high-fidelity results while achieving a computational speed-up of $\approx 2.4 \times 10^4$. We test our parametric SLM method for a two-parameter quarkyonic EOS against high-fidelity RK4 TOV calculations and find a computational speedup of $\approx 7.0 \times 10^4$. Hence, SLM is an efficient emulator for the numerous TOV evaluations required by multi-messenger astrophysical frameworks that infer constraints on the EOS. The ability of the SLM algorithm to learn a mapping between parameters of the EOS and subsequent neutron star properties also opens up potential extensions for assisting in computationally prohibitive uncertainty quantification (UQ) for any type of EOS. The source code for the methods employed in this work is openly available in a public GitHub repository for community modification and use.
\end{abstract}

\maketitle


\section{Introduction \label{sec:intro}}

The advent of multi-messenger astronomy, accelerated by LIGO's gravitational wave measurements~\cite{Abbott2017, LIGOScientific:2018cki, LIGOScientific:2020aai} and NICER's pulsar observations~\cite{Miller:2019cac, Miller:2021qha, Riley:2019yda, Riley:2021pdl, Choudhury:2024xbk}, has coincided perfectly with advances in Bayesian uncertainty quantification (UQ) in nuclear physics~\cite{Phillips:2020dmw, Melendez:2019izc} and the FRIB era of experiment~\cite{Brown:2024rml}. These three complementary approaches enable a refined understanding of the strong interaction, which governs matter from finite nuclei to dense, stellar compact objects, e.g., neutron stars. These intriguing astrophysical laboratories provide us with critical insights on extreme matter not found terrestrially. However, the lack of direct probes into the interior composition of neutron star cores indicates we must rely on the combination of information from neutron-star observations and binary mergers, nuclear experiment~\cite{Sorensen:2023zkk,MUSES:2023hyz}, and the construction of state-of-the-art nuclear potentials to infer the equation of state (EOS) of the star~\cite{Drischler:2021kxf, Lattimer:2021emm, Koehn:2024set}. This thermodynamic description, defined in terms of the state variables, i.e., pressure, number density, and energy density, is then propagated through the Tolman-Oppenheimer-Volkoff (TOV) equations~\cite{tolman1939, oppenheimer1939} to determine the corresponding mass and radius of the neutron star. Future gravitational wave (GW) measurements, as well as data from NICER and Cosmic Explorer, will provide more constraints on this mass-radius relationship, including other important properties, such as the tidal deformability~\cite{Thorne:1967, Hinderer:2007mb,Hinderer:2009ca, Postnikov:2010yn, Moustakidis:2017prq}, which will, in turn, allow us to place tighter constraints on the EOS.

This multi-messenger era of astrophysics has led to the need for more efficient calculations of the TOV equations due to their use in large-scale frameworks, e.g., LIGO's GW inference framework~\cite{Pang:2022rzc, KAGRA:2021duu, LIGOScientific:2017vwq}, where thousands of solutions can be required for reliable UQ of the mass-radius posterior. The precision era of nuclear physics~\cite{Drischler:2020yad, Drischler:2020hwi, Drischler:2020fvz, Semposki:2024vnp, Koehn:2024set, Hu:2021trw} has also led to the use of smaller-scale Bayesian frameworks that require sampling of the EOS and the computation of posteriors for neutron star properties~\cite{Reed:2024urq, Semposki:2024vnp, Koehn:2024set, Pang:2023dqj}, hence necessitating numerous solutions to the TOV equations. To perform these calculations efficiently, the computational burden of solving the TOV equations, especially when including tidal deformability, must be overcome.
An approach already employed in nuclear physics is the design and use of sophisticated emulators to reduce computational time and resources. Recent efforts in constructing model-intrusive, projection-based emulators have produced remarkable results in nuclear scattering~\cite{Drischler:2022ipa, Bonilla:2022rph, Maldonado:2024, Giuliani:2022yna, Odell:2023cun}, and model-extrusive emulators such as Gaussian processes (GPs) have been used successfully in nuclear astrophysics~\cite{Essick:2020flb, Semposki:2024vnp}.

In this work, we present a novel emulation strategy based upon the Dynamic Mode Decomposition (DMD) method used in analyses of time dynamics~\cite{Jonathan2014, Kutz:2016, brunton2021,2023Huhn}, supplemented by machine learning techniques. To differentiate this from standard DMD, we call our method Star Log-extended eMulation (SLM). Currently available DMD libraries~\cite{Demo2018} are incapable of handling the nonlinearity of the TOV equations; hence, this work establishes an emulator that overcomes this hurdle. We apply two SLM emulators to the TOV equations, including tidal deformability, for the first time.

This paper is organized as follows. In Sec.~\ref{sec:nsproperties} we review the TOV equations, including tidal deformability, and scale them for use with our high fidelity RK4 solver. In Sec.~\ref{sec:introdmd}, we give a general introduction to DMDs, which will be the foundation on which we build our SLM routines. We then detail the formalism, implementation and results of our initial SLM routine in Secs.~\ref{sec:slm}, which we apply to five tabular EOSs. We use these to benchmark our emulator against their corresponding high fidelity (HF) solutions. We then employ a parametric version of our emulator, called parametric SLM (pSLM), which we introduce and apply to a two-parameter quarkyonic EOS in Sec.~\ref{sec:pslm}. This emulator learns a mapping between EOS parameters and previously mentioned neutron star properties, a necessary feature for the Bayesian calibration of other phenomenological EOSs with freely varying parameters. We compare our SLM and pSLM results to low fidelity solvers, i.e., RK2 and forward Euler methods, in Sec.~\ref{sec:Compare}. Finally, we summarize our results and point to future applications of our work in Sec.~\ref{sec:summary}.

Our emulators, along with the code to produce the results in this paper, are publicly available for community use in our GitHub repository~\cite{DMDGitHub}.


\section{Neutron star properties \label{sec:nsproperties}}

Macroscopic properties of neutron stars, such as mass, radius, and tidal deformability are determined by solving the TOV equations simultaneously with appropriate differential equations corresponding to each additional desired quantity. The TOV equations, along with the tidal deformability equation, are given by~\cite{tolman1939, lindblom1992, Hinderer:2007mb, Hinderer:2009ca, Postnikov:2010yn},
\begin{align}
    \frac{dP}{dr} &= -\frac{G}{c^2}\left[\epsilon(r) + P(r)\right] \frac{m(r) + 4\pi r^3 P(r)/c^2}{r\left[r-2Gm(r)/c^2\right]}, \\
    \frac{dm}{dr} &= 4\pi r^2 \frac{\epsilon(r)}{c^2}, \\
    \frac{dy}{dr} &= -\frac{y(r)^2}{r} - \frac{F(r) y(r)}{r}  - \frac{Q(r)}{r},
\end{align}
where,
\begin{align}
    F(r) = \frac{1 - 4\pi G r^2 \left[\epsilon(r) - P(r) \right]/c^4}{1 - 2\frac{Gm(r)}{rc^2}}, 
\end{align}
and
\begin{equation}
\begin{aligned}
   Q(r) &= \frac{4\pi G r^2/c^4}{1- \frac{2Gm(r)}{rc^2}} \left( 5 \epsilon(r) + 9 p(r) + \frac{\epsilon(r) + p(r)}{c_s(r)^2}c^2 - \right. \\
   &\left. \frac{6c^4}{4\pi r^2 G} \right)
   - 4 \left(\frac{G\left[m(r)/(rc^2) + 4\pi r^2 p(r)/c^4\right]}{1 - 2 Gm(r)/(rc^2)} \right)^2.
\end{aligned}    
\end{equation}

The dimensionless versions of these equations can be obtained by scaling them using the parameters in Ref.~\cite{Piekarewicz2017}. We scale the differential equation for tidal deformability in a similar fashion (see Appendix~\ref{TOVdetails} for details). The resulting equations are
\begin{align}
    \label{eq:TOVp}
    \frac{dp}{dx} & = -\frac{1}{2} \frac{\left[\varepsilon(x) + p(x)\right]\left[m(x) + 3x^3 p(x)\right]}{x^2\left(1 - m(x)/x\right)}, \\
    \label{eq:TOVm}
    \frac{dm}{dx} &= 3x^2 \varepsilon(x), \\
    \label{eq:tidaly}
    \frac{dy}{dx} &= -\frac{y(x)^2}{x} - \frac{F(x)y(x)}{x} - \frac{Q(x)}{x}, 
\end{align}
with
\begin{equation}
    F(x) = \left[1 - \tfrac{3}{2} x^2 (\varepsilon(x) - p(x))\right][1-m(x)/x]^{-1},
\end{equation}
and 
\begin{align}
    Q(x) &= \frac{\tfrac{3}{2} x^2}{\left(1 - \frac{m(x)}{x} \right)} \Bigg[5\varepsilon(x) + 9p(x) + \frac{\varepsilon(x) + p(x)}{c_s^2(x)}c^2 \nonumber \\
    &~~~~~- \frac{4}{x^2} \Bigg] - \left(\frac{m(x)/x + 3x^2 p(x)}{\left(1 - \frac{m(x)}{x} \right)} \right)^2,
\end{align}
where $p(x)$, $\varepsilon(x)$, $m(x)$ and $c_s^{2}(x)$ are the scaled dimensionless pressure, energy density, mass and squared speed of sound functions respectively, while $y(x)$ is the scaled solution to the tidal equation. The dense matter EOS is an essential input to this system through $p(x)$ and $\varepsilon(x)$.

We solve~\cref{eq:TOVp,eq:TOVm,eq:tidaly} as initial value problems for a given central pressure $p_c$, mass $m = 0$ and $y = 2.0$ (from Ref.~\cite{Postnikov:2010yn}) at the center of the star. The radius and mass of the star are determined by the condition $p(x) \rightarrow 0$ at the edge of the star. In this work we have employed a high-fidelity (HF) fourth order Runge-Kutta (RK4) solver. We also construct the dimensionless coefficient, $k_2(R)$, known as the Love number~\cite{Postnikov:2010yn}, which is computed at the radius $R$ of the star. $k_{2}(R)$ is computed as
\begin{align}
    \label{eq:lovenumber}
    k_2(R) &= \frac{8\beta^5}{5} \left(1 - 2\beta \right)^2 \left[2 + 2\beta(y_R -1) -y_R \right] \nonumber \\
    &\times \left\{ 2\beta(6 - 3y_R + 3\beta(5y_R -8)) \right. \nonumber \\
    &\left.+ 4\beta^3\left[13 - 11y_R + \beta(3y_R - 2) + 2\beta^2(1+ y_R)\right] \right. \nonumber \\
    &\left. + 3(1-2\beta)^2[2 - y_R + 2\beta(y_R - 1)]\log(1-2\beta) \right\}^{-1},
\end{align}
where $\beta = GM(R)/R$ is the compactness of the star, and $y_R = y(r=R)$. Here, we use the unscaled solutions of the TOV equations.

We have chosen input EOSs to the TOV equations that can obtain a maximum mass of greater than $2M_{\odot}$, possess different sets of underlying parameters, and enable us to test our algorithm on a variety of masses and radii, as well as various structures of the Love number curve $k_{2}(R)$. Our HF solutions for their maximum masses and corresponding radii are presented in Table~\ref{tab:eos_table}, and are consistent with the results in the compOSE database~\cite{CompOSECoreTeam:2022ddl} and from the original publications~\cite{Chen:2014mza}. In Secs.~\ref{sec:slmresults} and Sec.~\ref{sec:pslmresults} we compare these HF solutions to our emulated results.


\section{Dynamic Mode Decomposition}
\label{sec:introdmd}

\begin{algorithm*}[t]     
    \caption{SLM. We closely follow Refs.~\cite{Kutz:2016, 2023Huhn} for the foundation of the SLM algorithm (see Alg.~\ref{alg:slm}). Italicized portions of the algorithm are direct extensions to Refs.~\cite{Kutz:2016, 2023Huhn}.} \label{alg:slm}
    1: Solve Eq. (\ref{eq:slm4}) and collect snapshots $[\vec{\xi}_i]^{m}_{i=0}$ \textit{where} $i$ \textit{are the data indices.} \\
    2: \textit{Compute the logarithm of these snapshots as} $[\log\vec{\xi}(r_i)]^{m}_{i=0}$. \\
    3: \textit{Extend the snapshots by adding the Hadamard product of the arrays, i.e.} $\bm{\prod}_{i,j} \log\vec{\xi}_i\log \vec{\xi}_j$. \\
    4: Organize the snapshots in $\mathbf{X}$ and $\mathbf{X}'$ data matrices. \\
    5: Perform SVD of $\mathbf{X}$: $\mathbf{X} = \mathbf{U\Sigma V^T}$. \\
    6: Keep $s$ modes to desired accuracy and compute the reduced Koopman operator $\mathbf{A}_s \equiv \mathbf{U}^T_s\mathbf{A}\mathbf{U}_s = \mathbf{U}^T_s\mathbf{X}'\mathbf{V}_s\mathbf{\Sigma}_s^{-1}$. \\
    7: Perform eigen-decomposition of $\mathbf{A}_s$ to obtain the reduced eigen-modes: $\mathbf{A}_s\mathbf{W} = \mathbf{\Lambda W}$. \\
    8: Produce the full-state modes $\mathbf{\Phi} = \mathbf{U}_s \mathbf{W}$. \\
    9: Reconstruct the coupled, \textit{nonlinear} ODE solutions $\vec{\xi}(r)$ using the first $n$ modes with Eq.~\eqref{eq:dmdreducedsolns} and \textit{exponentiate the solutions}. \\
\end{algorithm*}

Dynamic mode decomposition (DMD)~\cite{Jonathan2014, Kutz:2016}, a data-driven technique, is a model-extrusive emulator and has been successful in emulating dynamical systems in the areas of fluid dynamics, neuroscience and financial systems~\cite{BRUNTON20161, Mann01112016, Jiaqing2018}. Fundamentally, the method can be viewed as a blend of spatial dimensionality reduction techniques, such as proper orthogonal decomposition (POD), and Fourier transforms in the time domain. In this section, we follow Ref.~\cite{Kutz:2016} closely.

Consider a set of non-autonomous coupled differential equations
\begin{equation}
    \frac{\p \vec{\xi}}{\p t} = \mathcal{F}(\vec{\xi}(t), t), \quad \vec{\xi}(t) \in \mathbb{R}^{n},
\end{equation}
where $\mathcal{F}$ is a nonlinear operator that describes the dynamics of the governing equations of interest $\vec{\xi}(t)$, and where $\vec{\xi}(t)$ is the system's state at some given time $t$. Finding the exact form of $\mathcal{F}$ can be prohibitively difficult, so to mitigate this issue we employ the approximately locally linear characteristic of these systems, discretize the phase space in terms of equally spaced time steps $\Delta t$ and write
\begin{equation}
    \label{eq:locallinearapprox}
    \frac{d\vec{\xi}}{dt} = \mathbf{\mathcal{A}}\vec{\xi},
\end{equation}
where $\mathbf{\mathcal{A}}$ is the locally linear approximation of the dynamics originally described by $\mathcal{F}$. Hence, it locally governs the evolution of the system, even if the true dynamics are nonlinear. However, in practice we need to discretize this in terms of time steps $\Delta t$, so we do this by writing a discretized version of $\mathcal{A}$: $\mathbf{A} = \exp(\mathbf{\mathcal{A}}\Delta t)$. Eq.~\eqref{eq:locallinearapprox} hence becomes
\begin{equation}
    \label{eq:realdmdequation}
    \frac{d\vec{\xi}}{dt} = \mathbf{A}\vec{\xi}.
\end{equation}
To determine $\mathbf{A}$, we will construct matrices of snapshots of our system at different time steps. These are generally denoted as
\begin{equation}
    \label{eq:snapshotmatrices}
  \mathbf{X} = 
  \begin{bmatrix}
      | & | & & | \\
      \xi_{1} & \xi_{2} & \ldots & \xi_{m-1} \\
      | & | & & | \\
  \end{bmatrix}; 
  \qquad 
  \mathbf{X}' = 
  \begin{bmatrix}
      | & | & & | \\
      \xi_{2} & \xi_{3} & \ldots & \xi_{m} \\
      | & | & & | \\
  \end{bmatrix}.
\end{equation}

The locally linear approximation is devised from here by writing $\mathbf{X}' \approx \mathbf{A} \mathbf{X}$. To find the best-fit value of $\mathbf{A}$, we can then write $\mathbf{A} = \mathbf{X}'\mathbf{X}^\dagger$ using the Moore-Penrose pseudoinverse. To save computational time and effort, and to ensure our results are numerically stable, we then find the reduced rank evolution operator, which we denote as $\mathbf{A}_{s}$; this is done by projecting the snapshot matrix $\mathbf{X}$ into a low-dimensional subspace via the Singular Value Decomposition (SVD),
\begin{equation}
    \mathbf{X} = \mathbf{U} \mathbf{\Sigma} \mathbf{V}^*,
\end{equation}
where $\mathbf{U} \in \mathbb{C}^{m \times m}$, $\mathbf{\Sigma} \in \mathbb{C}^{m \times n}$, and $\mathbf{V} \in \mathbb{C}^{n \times n}$. We pick $s$ singular values (also called modes) that approximate $\mathbf{A}$ to a desired accuracy. This can be achieved by
\begin{equation} \label{eq:dmd6}
    s = \mathrm{argmin}_j \frac{\Sigma_{i=1}^{i=j}\sigma_i}{\Sigma_{i=1}^{i=k}\sigma_i} , \eta
\end{equation}
where $\sigma$ are the $j$ singular values chosen from the SVD, $k$ is the total number of singular values, and $\eta$ is a value between 0 and 1 that we must choose for our problem~\cite{2023Huhn}.

Hence, we now have $\mathbf{U}_s \in \mathbb{C}^{m \times s}$, $\mathbf{\Sigma}_s \in \mathbb{C}^{s \times s}$, and $\mathbf{V}_s \in \mathbb{C}^{n \times s}$. We use these reduced matrices to express $\mathbf{A}_s$ via the pseudoinverse $(\mathbf{X}^{\dagger})$ as 
\begin{equation}
    \mathbf{X}^{\dagger} \approx \mathbf{V}_s \mathbf{\Sigma}_s^{-1} \mathbf{U}_s^{*}
\end{equation} to obtain
\begin{equation}
    \mathbf{A}_s = \mathbf{X}'\mathbf{V}_s\mathbf{\Sigma}_s^{-1}\mathbf{U}_s^*.
\end{equation}

Now, to derive $\mathbf{A}_s$ in the projected low-dimensional subspace, we take the projected matrix using
\begin{equation}
    \mathbf{A}_s = \mathbf{U}^{*}_s\mathbf{A}_s\mathbf{U}_s = \mathbf{U}_s^{*}\mathbf{X}_s'\mathbf{V}_s\mathbf{\Sigma}_s^{-1}.
\end{equation}
From here, we can compute the eigendecomposition of $\mathbf{A}_s$ using
\begin{equation}
    \mathbf{A}_s\mathbf{W}_s = \mathbf{W}_s \mathbf{\Lambda}_s,
\end{equation}
where $\mathbf{\Lambda}_s$ is the diagonal matrix of the eigenvalues of $\mathbf{A}_s$ and $\mathbf{W}_s$ is the matrix of corresponding eigenvectors. 

We can then construct the eigenvectors $\mathbf{\Phi}$ of the solutions $\vec{\xi}(t)$ through the ``projected DMD modes", which are found through $\mathbf{\Phi} = \mathbf{U}_s\mathbf{W}_s$. We also define $\omega_{k} = \ln{(\lambda_{k})}/\Delta t$ and obtain $\vec{b}$ from the pseudoinverse $\vec{b} = \mathbf{\Phi}^{\dagger} \vec{\xi}$. Finally, we can construct the equation of solutions to our system, which are
\begin{equation}
    \label{eq:dmdreducedsolns}
    \vec{\xi}(t) = \sum_{k=1}^{r} \vec{\phi}_{k} \exp{(\omega_{k}t)} b_{k} \equiv \bm{\Phi}\exp(\bm{\Omega} t)\vec{b}.
\end{equation}
for all future times. This equation-free structure makes DMDs very efficient; they are constrained only by the size of the rank $s$ chosen. The emulators in the following two sections are built upon this standard DMD, but are designed to obtain better results for our nonlinear TOV system. 


\section{Star Log-extended eMulation \label{sec:slm}}
Here, we use the foundational aspects of DMDs to build our emulator for the TOV equations. We discuss the formalism of this emulator in Sec.~\ref{sec:slmformalism}, and move to our results for a selection of tabular EOSs in Sec.~\ref{sec:slmresults}.

\subsection{Formalism} \label{sec:slmformalism}

DMD-based emulators are generally applicable to problems that have polynomial nonlinearities or can be expressed as an eigenvalue problem~\cite{Drischler:2022ipa, Odell:2023cun, Bonilla:2022rph, Melendez:2022kid}, and are used extensively for the emulation of various time-dependent systems~\cite{Korda2018, baddoo2021, 2023Huhn, velegar2024}. Recent works applied DMDs to nonlinear problems~\cite{2023Huhn} including a time-independent scenario~\cite{sharak2022}; however, the DMD algorithms presented there suffer from the inability to capture the more complex nonlinear dynamics of the coupled TOV system. To overcome this limitation, we extend the DMD method to one that can handle the TOV equations by building a logarithmic structure into the DMD modes.

\begin{figure*}[t]
\centerline{\includegraphics[width=\linewidth]{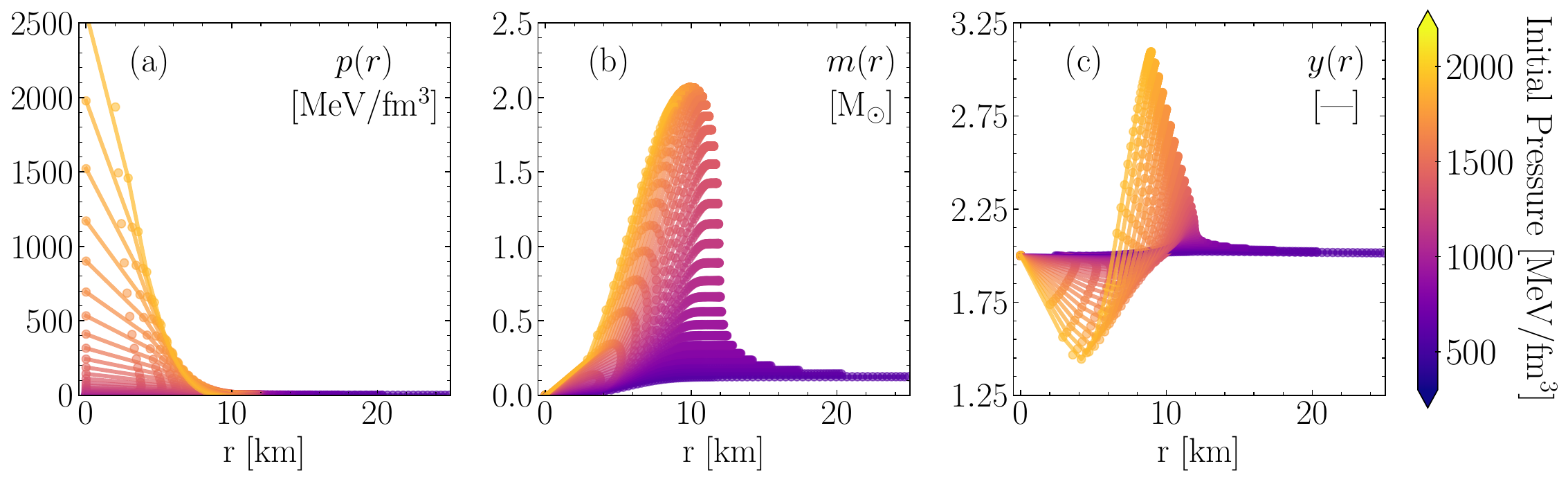}}
\caption{High fidelity solutions to the TOV and tidal deformability coupled differential equations (dotted curves) for the SLy4 tabular EOS~\cite{CompOSECoreTeam:2022ddl} at various initial pressures, and the corresponding SLM predictions (solid curves). These results highlight the ability of the SLM routine to capture the underlying structure of the nonlinear differential equations.}
\label{fig:tovSoln}
\end{figure*}

Thus, in our case, we solve a non-autonomous system of equations, which can be written as
\begin{equation} \label{eq:slm4}
    \frac{d\vec{\xi}}{dr} = \mathcal{F}(\vec{\xi}(\vec{\alpha}, r), r),
\end{equation}
where $\mathcal{F}$ is a nonlinear operator that describes the governing equation of interest by discretizing the phase space, i.e., taking snapshots at discrete values of $r$. Here $r$ is the independent variable, while $\vec{\alpha}$ consists of other dependent variables such as $\varepsilon$.
We therefore wish to find
\begin{equation} \label{eq:slm_b}
    \frac{d\vec{\xi}}{dr} = \mathbf{A}\vec{\xi},
\end{equation}
where the linear operator $\mathbf{A}$ approximates the dynamics of the system as in Sec.~\ref{sec:introdmd}. This operator is learned in a data-driven fashion. 

In the DMD procedure, we replace the independent variable $t$ with the linear index $\ell$ of the data in the training set, and include the radius $r$ in the set of snapshots directly. Since the linear index is linearly spaced and monotonically increasing, this helps us handle the nonlinearity in the spacing of the training set.

Our snapshot matrices $\mathbf{X}$ and $\mathbf{X}'$ now can contain either the (unscaled) training data from the solutions of the TOV equations, i.e., $m(r)$, $p(r)$, $y(r)$, and $r$, or the final curves corresponding to $M(R)$, radius $R$, central pressure $P_{\mathrm{c}}(R)$ and $k_2(R)$. In the first case, the SLM routine will train on the profile curves from the various initial pressures used to generate the TOV results for a given EOS. In the second case, the SLM routine instead trains on the final curves that are obtained from finding the points on the profile curve that correspond to the radius $R$. Results for both of these cases are shown in the next section.

Once we choose which training data to use, we then take the logarithm of these snapshots, i.e., $\vec{z} = \log \{\vec{\xi}_i\}_{i=0}^m \in \mathbb{R}^{n\times m}$, where $i$ indexes the $m$ total snapshots. Thus, in the first case, $\vec{z}$ contains solutions corresponding to $\log(y(r))$, $\log(m(r))$, $\log(p(r))$, and $\log(r)$. We then employ the formalism of extended DMDs in Ref.~\cite{williams2015}, which is a machine learning technique, to modify the snapshots further to make them suitable for our calculations, i.e., including not only the snapshot vectors themselves in the snapshot matrices $\mathbf{X}$ and $\mathbf{X}'$, but also taking the Hadamard product of the snapshot vectors. This approach significantly enhances reconstruction and prediction capabilities of the data. 

The snapshots $\vec{z}$ and their Hadamard products are arranged in the matrices $\mathbf{X}$ and $\mathbf{X}' \in \mathbb{R}^{n\times m(m-1)}$, which are now defined as
\begin{align} \label{eq:dmd5}
    \mathbf{X} &= \begin{bmatrix} \vec{z}_{1} & \ldots & \vec{z}_{m-1} &\vec{z}_1^{\,2} & \vec{z}_1\vec{z}_2 &\ldots &\vec{z}_i\vec{z}_j &\ldots &\vec{z}_{m-1}^{\,2}\end{bmatrix}, \nonumber\\  \mathbf{X'} &= \begin{bmatrix} \vec{z}_{2} & \ldots & \vec{z}_{m} &\vec{z}_2^{\,2} & \vec{z}_2\vec{z}_3 &\ldots &\vec{z}_i\vec{z}_j &\ldots &\vec{z}_{m}^{\,2}\end{bmatrix}.
\end{align}
We then follow the DMD procedure outlined in Sec.~\ref{sec:introdmd}, and included in algorithm~\ref{alg:slm}, using these snapshot matrices. For our use of SLM with the TOV equations, we chose $\eta = 0.9999$ in Eq.~\eqref{eq:dmd6} to determine the number of modes for the SLM routine.

\begin{table*}[ht!]
    \centering
    \begin{tabular}{|c|c|c|c|c|c|c|c|c|}
        \hline
        & \multicolumn{2}{c|}{HF} & \multicolumn{2}{c|}{SLM} & \multicolumn{2}{c|}{Rel. Error (\%)} & \multicolumn{2}{c|}{Time (s)} \\
        \hline
         \multirow{2}{*}{EOS} & Max. & Radius & Max. & Radius & Max. & \multirow{2}{*}{Radius} & \multirow{2}{*}{HF} & \multirow{2}{*}{SLM}  \\
         & Mass [$M_\odot$] & [km] & Mass [$M_\odot$] & [km] & Mass & & & \\
         \hline
         \multicolumn{9}{|c|}{Tabular EOS} \\
         \hline
         SLy4 & 2.067 & 10.033 & 2.066 & 10.031 & 0.04 & 0.01 &
         10.711 & $3.841\times 10^{-4}$ \\
         APR & 2.193 & 9.979 & 2.193 & 9.981 & 0.01 & 0.02 & 
         10.856 & $4.580\times 10^{-4}$ \\
         FSU Garnet & 2.066 & 11.601 & 2.066 & 11.598 & 0.00 & 0.02 &
         10.934 & $4.520\times 10^{-4}$ \\
         BL  &  2.083 & 10.341 & 2.083 & 10.345 & 0.01 & 0.04 &
         10.767 & $5.159\times 10^{-4}$ \\
         DS-CMF-5 & 2.023 & 11.802 & 2.032 & 11.597 & 0.46 & 1.73 &
         10.827 & $4.659\times 10^{-4}$ \\
         \hline
         \multicolumn{9}{|c|}{Quarkyonic EOS} \\
         \hline
         $\Lambda$=300.00, $\kappa$=0.26 & 2.666 & 12.614  & 2.658 & 12.642 & 0.30 & 0.22 &
         11.718 & $1.600\times 10^{-4}$ \\
         $\Lambda$=352.63, $\kappa$=0.12 & 2.329 & 10.671 & 2.322 & 10.632 & 0.29 & 0.36 &
         11.421 & $1.643\times 10^{-4}$ \\
         $\Lambda$=384.21, $\kappa$=0.23 & 2.476 & 11.517 & 2.454 & 11.521 & 0.88 & 0.04 &
         11.323 & $1.559\times 10^{-4}$ \\
         $\Lambda$=436.84, $\kappa$=0.14 & 2.887 & 14.532 & 2.891 & 14.580 & 0.14 & 0.33 &
         11.238 & $1.540\times 10^{-4}$ \\
         $\Lambda$=500.00, $\kappa$=0.28 & 2.224 & 10.177 & 2.225 & 10.172 & 0.06 & 0.05 &
         11.225 & $1.788\times 10^{-4}$ \\
         \hline
    \end{tabular}
    \caption{Comparisons of the HF and emulated maximum mass and corresponding radii for various EOSs~\cite{Bombaci:2018ksa,Chen:2014mza,Chabanat:1997,Dexheimer:2021, Dexheimer:2008ax,Akmal:1998} selected from the compOSE database~\cite{CompOSECoreTeam:2022ddl}, as well as the same quantities computed for the quarkyonic EOS~\cite{McLerran:2018hbz} with various choices of parameters $\Lambda$ and $\kappa$ are shown. The tabular EOSs are emulated using the SLM algorithm (see Fig.~\ref{fig:SLM_Data}), whereas the quarkyonic EOS is emulated through the pSLM algorithm (see Fig.~\ref{fig:parametricDMDs}). The remarkable efficiency achieved is seen in the large difference between the time spent computing the HF and emulated solutions (last two columns), leading to an average speed-up of $4.7\times 10^{4}$.}
    \label{tab:eos_table}
\end{table*}

\subsection{Results} \label{sec:slmresults}

As a test of our algorithm, we first emulate the solutions of the $m(r)$, $p(r)$, and $y(r)$ profile curves of a selected EOS. For this task, we utilize the well-known SLy4 EOS as input to the TOV equations for our HF calculations. Figure~\ref{fig:tovSoln} shows the solutions to Eqs.~\eqref{eq:TOVp}-\eqref{eq:tidaly} as functions of radius $r$ and the output from the SLM algorithm for each quantity, respectively. The emulator is able to recover the solutions from the HF solver, allowing for accurate predictions between HF solutions. Hence, SLM only needs a few solutions of each quantity to reliably reconstruct the HF results.\footnote{To construct a mass-radius curve from these SLM solutions, we can impose the condition $p = 0$ at $r = R$ and collect the values of the mass $(M = m(r=R))$ (and $y(r=R)$ for the tidal deformability if desired). Hence, SLM can be used to emulate the profiles of the EOS and the result can be propagated to the final $M-R$ curve.}

\begin{figure}[t]
\centerline{\includegraphics[width=0.92\columnwidth]{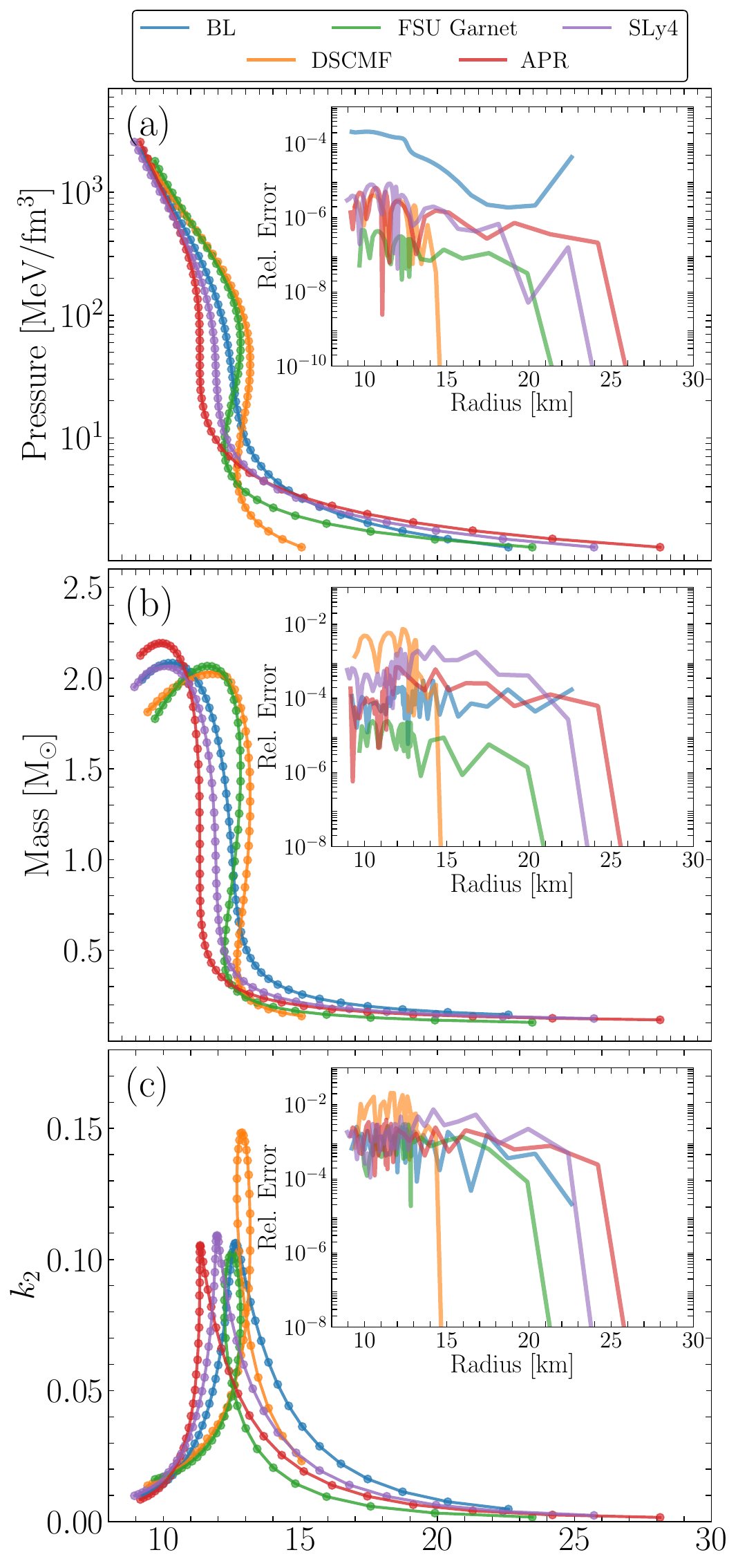}}
\caption{(a) Central pressure, (b) mass, and (c) dimensionless Love number $k_2$ as functions of radius. These properties are calculated for five tabular EOSs~\cite{Bombaci:2018ksa,Chen:2014mza,Chabanat:1997,Dexheimer:2021, Dexheimer:2008ax,Akmal:1998}. The dots correspond to the HF results, and the solid lines indicate the SLM predictions. The inset shows the relative error between the HF and SLM results.
}
\label{fig:SLM_Data}
\end{figure}

To apply the SLM algorithm directly to the HF calculations of the $M-R$, $P_{\textrm{c}}-R$, and $k_2-R$ curves, we choose five realistic equations of state that can produce a maximum mass of $\geqslant 2 M_{\odot}$~\cite{Bombaci:2018ksa,Chen:2014mza,Chabanat:1997,Dexheimer:2021, Dexheimer:2008ax,Akmal:1998}. These EOSs have different sets of underlying parameters. For instance, the APR is a non-relativistic EOS employing the Urbana-Argonne potential, while FSU Garnet is a relativistic mean-field EOS. Figure~\ref{fig:SLM_Data} shows our SLM routine applied to these HF calculations using 50 total snapshots. The respective relative errors between the HF and emulated results were found to be $\leqslant 3 \times 10^{-3}$. These values are shown in the insets in Fig.~\ref{fig:SLM_Data}. With an average speed of about $4.6\times 10^{-4}$ seconds, compared to the HF average speed of $\approx 11$ seconds, the SLM routine is shown to be remarkably computationally efficient. This proof-of-principle calculation indicates that SLM can be used to accurately recover the structure of each curve through prediction between HF solutions corresponding to a given EOS. 

Additionally, Table~\ref{tab:eos_table} shows the numerical results for the maximum masses and corresponding radii calculated from the HF RK4 solver for these EOSs, compared with the same quantities using the SLM algorithm, and the subsequent percent error between the HF and emulated calculations. All errors are $\leq 2\%$ for the maximum mass and the corresponding radii.

\begin{figure}[t]
\centerline{\includegraphics[width=0.92\linewidth]{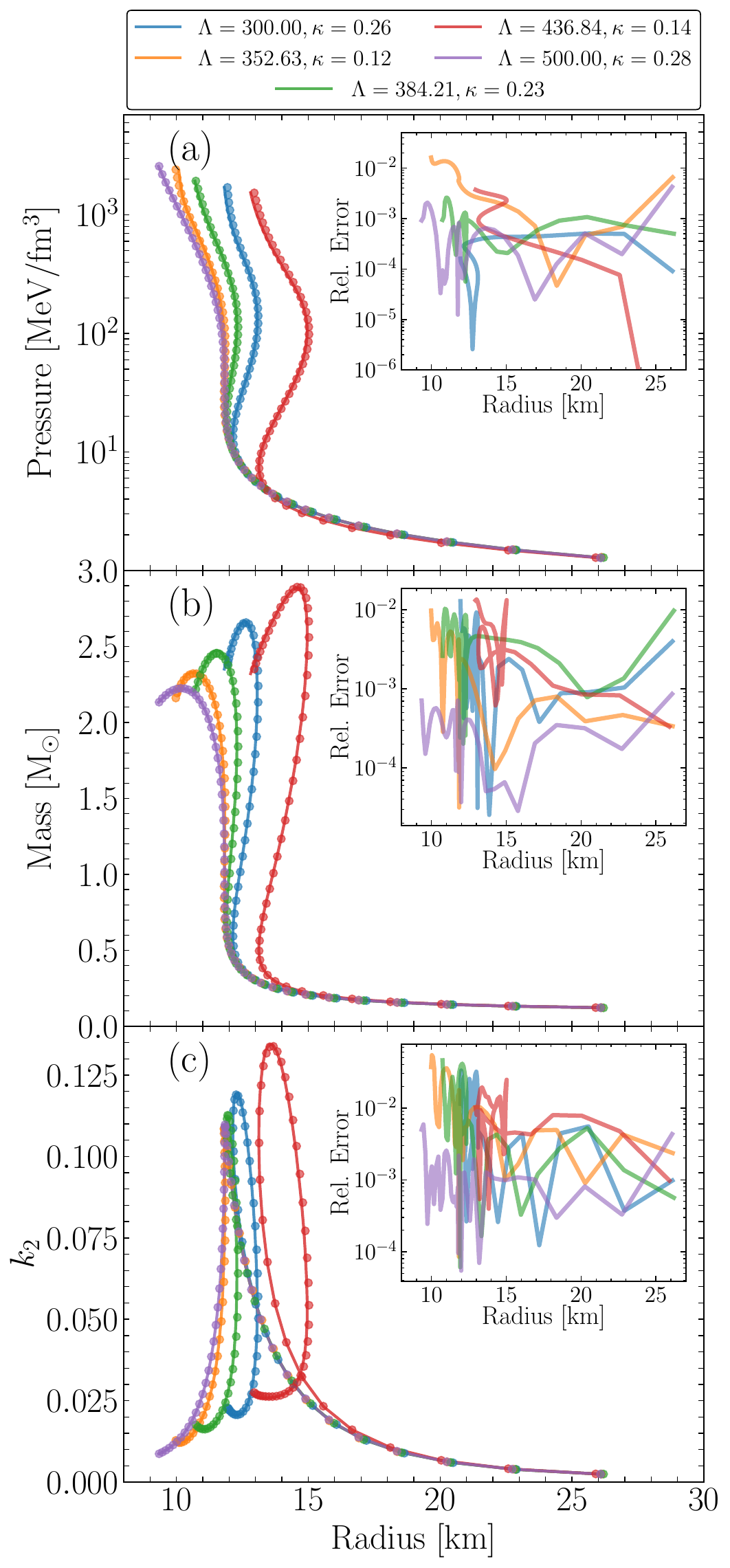}}
\caption{(a) Central pressure, (b) mass, and (c) dimensionless Love number $k_{2}$ as function of radius for the quarkyonic EOS~\cite{McLerran:2018hbz} using pSLM, with varying parameters $\Lambda$ and $\kappa$. The result shown here is the prediction for a given test parameter set that has not been used to train the emulator. The dots (solid lines) indicate SLM (HF) results.
}
\label{fig:parametricDMDs}
\end{figure}


\section{Parametric SLM \label{sec:pslm}}

We now turn to the parametric version of our SLM emulator, which we call parametric SLM (pSLM). We discuss the modifications to our SLM algorithm for the parametric case in Sec.~\ref{sec:pslmformalism}, and then show our implementation and results in Sec.~\ref{sec:pslmresults}.

\begin{algorithm*}[t]    
    \caption{
    Parametric SLM. As with Alg.~\ref{alg:slm}, the italicized portions of this algorithm are direct extensions to Refs.~\cite{Kutz:2016, 2023Huhn}.
    }\label{algo:pSLM}
        1: Solve Eq.~(\ref{eq:pdmd}) for all training parameters $\vec{\mu}_j$ and collect $N_s$ snapshots $\mathbf{X}_{\vec{\mu}_j} = [\vec{\xi}(r_i; \vec{\mu}_j)]^{m}_{i=0}$ for $1 \leq j \leq N_s$. \\
        2: \textit{Compute the logarithm of these snapshots} $[\log\vec{\xi}(r_i; \vec{\mu}_j)]^{m}_{i=0}$. \\
        3: \textit{Extend the snapshots by adding the Hadamard product of the arrays, i.e. }$\left[\bm{\prod}_{i,k} \log\vec{\xi}_i\log \vec{\xi}_k\right]_{j},~1 \leq j \leq J$. \\
        4: Organize the snapshots in $\mathbf{X}_j^+$ and $\mathbf{X}_j^-$ data matrices, $1\leq j \leq J$. \\
        5: Perform SVD of $\mathbf{X}_j^-$: $\mathbf{X}_j^- = \mathbf{U}_j\mathbf{\Sigma}_j \mathbf{V}_j^T $, $1 \leq j \leq J$. \\
        6: Retain $s$ modes and compute the reduced Koopman operator $\mathbf{A}_{js} \equiv \mathbf{U}^T_{js}\mathbf{AU}_{js} = \mathbf{U}^T_{js}\mathbf{X}^+_j \mathbf{V}_{js}\mathbf{\Sigma}_{js}^{-1}$, $1\leq j \leq J$. \\
        7: Perform the eigen-decomposition of $\mathbf{A}_{js}$ to obtain the reduced eigen-modes as described in Eq.~(\ref{eq:eigen-A}). \\
        8: Interpolate SVD-modes $\mathbf{U}_j$, eigen-modes $\mathbf{W}_{js}$, and eigen-values $\mathbf{\Lambda}_{js}$ to obtain $\mathbf{U}_{\theta s}$, $\mathbf{W}_{\theta s}$, and $\mathbf{\Lambda}_{\theta s}$, respectively, \textit{using the B-GRIM interpolation method}. \\
        9: Construct the modes $\mathbf{\Phi}_{\theta} = \mathbf{U}_{\theta s}\mathbf{W}_{\theta s}$. \\
        10: Recover the initial coefficients $\vec{b}_{\theta}$  by interpolating $\vec{b}_j, 1 \leq j \leq J$. \\
        11: Reconstruct the coupled set of solutions \textit{and exponentiate} to obtain $\vec{\xi}^{\,\theta}(r)$. \\
\end{algorithm*}

\subsection{Formalism} \label{sec:pslmformalism}

The parametric SLM (pSLM) method is based upon the reduced Eigen-Pair Interpolation method (rEPI) from Ref.~\cite{2023Huhn}, using our modified snapshots (see Sec.~\ref{sec:slmformalism}). However, instead of using a point-wise Lagrange interpolation as in Ref.~\cite{2023Huhn}, we use a recently developed technique based on the Greedy Recombination Interpolation Method (GRIM) for Banach spaces (B-GRIM)~\cite{lyons2024} to interpolate more accurately within the parameter space. 

The pSLM method expands upon the general algorithm for SLM presented in the previous section, in order to determine the evolution of the system under parametric dependence. The governing equations in Eq.~(\ref{eq:slm4}) may involve specific parameters $\mu_i \in \vec{\mu}, i =1, 2, \ldots, n$ allowing them to be reformulated in a parametric form 
\begin{equation} \label{eq:pdmd}
    \frac{\p\vec{\xi}^{\,\vec{\mu}}}{\p r} = \mathcal{F}(\vec{\xi}^{\,\vec{\mu}}(\vec{\alpha}, r; \vec{\mu}), r; \vec{\mu}),
\end{equation}
where $\vec{\alpha}$ is a vector of other variables that $\vec{\xi}$ depends on.

After discretizing the phase space, this leads to a system of coupled parametric ordinary differential equations
\begin{equation} \label{eq:pdmd2}
    \frac{d\vec{\xi}^{\,\vec{\mu}}}{dr} = \mathcal{A}\vec{\xi}^{\,\vec{\mu}}(\vec{\alpha}, r; \vec{\mu}).
\end{equation}
The goal here is to replace the nonlinear operator $\mathcal{A}(\cdot)$ with a linear surrogate approximation
\begin{equation} \label{eq:app_pdmd3}
    \frac{d\vec{\xi}^{\,\vec{\mu}}}{dr} = \mathbf{A}(\vec{\mu})\vec{\xi}^{\,\vec{\mu}},
\end{equation}
where ${\vec{\mu}}$ is a vector of parameters. For a given vector of parameters $\mu_j\in \vec{\mu},\, j\in \{1, 2, \ldots, n\}$, HF solutions are again calculated at a few snapshots. Individual classical SLMs are subsequently performed upon these snapshots. 

We would like to use our algorithm to predict at test sets of parameters; we denote these as $\vec{\theta}$. The B-GRIM method is employed to identify the nearest neighbor of the test parameter $\theta$ from the training set of snapshots. This enables a targeted interpolation of the eigen-pairs associated with $\mathbf{A}$. Using this approach, we compute SVD decompositions of $J$ data sets by setting the accuracy using Eq.~\eqref{eq:dmd6} and select the highest of the $J$ ranks ($s = \min_{j\in J}(s_j)$), yielding the reduced left- and right-singular matrices $\mathbf{U}_{js}$ and $\mathbf{V}_{js}^T$ along with the corresponding reduced singular value matrix $\mathbf\Sigma_{js}$. The individual reduced Koopman operators, $\mathbf{A}_{js}$, for each of the $J$ cases are constructed as
\begin{equation}
    \mathbf{A}_{js} = \mathbf{U}_{js}^T\mathbf{X}_j'\mathbf{V}_{js}\mathbf{\Sigma}_{js}^{-1},
\end{equation}
from which the reduced eigen-pairs are obtained as
\begin{equation}
    \mathbf{A}_{js}\mathbf{W}_{js} = \mathbf{\Lambda}_{js}\mathbf{W}_{js}.
\label{eq:eigen-A}
\end{equation}
The desired test case eigen-pair $(\mathbf{\Lambda}_{\theta s}, \mathbf{W}_{\theta s})$ and $\mathbf{U}_{\theta s}$ and $\vec{b}_{\theta}$ are then determined via interpolation using B-GRIM, leveraging the $J$ eigen-pairs $(\mathbf{\Lambda}_{js}, \mathbf{W}_{js})$.\footnote{B-GRIM is optimal for high-dimensional spaces where residual minimization in Banach spaces is effective. This is especially true when dealing with noisy data or requiring iterative improvement of the solution, whereas Lagrange interpolation is unstable in such cases (see appendix~\ref{app:BGRIM} for more details).} The projected modes are subsequently calculated as $\mathbf{\Phi}_{\theta} = \mathbf{U}_{\theta s}\mathbf{W}_{\theta s}$. Finally we obtain the solutions for the test parameter set $\vec{\theta}$ using Eq.~(\ref{eq:dmdreducedsolns}) and keeping $s$ modes. The corresponding full algorithm for pSLM is given in Alg.~\ref{algo:pSLM}.


\subsection{Results} \label{sec:pslmresults}

To test our ability to generate solutions to parametric EOSs at various parameter sets, we use the quarkyonic EOS~\cite{McLerran:2018hbz} to test the pSLM algorithm. This EOS assumes that both nucleons and quarks are quasi-particles and that quarks reside in the Fermi sea, while, at a sufficiently large Fermi energy, nucleons appear as correlations at the edge of the Fermi surface. In momentum space, this corresponds to a sphere of quarks surrounded by a shell of baryons of width $\Delta$. To find this width, we use the expression
\begin{equation}
    \Delta = \frac{\Lambda^3}{k_{FB}^2} + \kappa \frac{\Lambda}{N_c^2},
\end{equation}
where $N_c$ is the number of quark colors and $k_{FB}$ is the Fermi momentum of the baryons. The two unknown parameters above, $\Lambda$ and $\kappa$, must be set to obtain the EOS. Hence, we vary these two parameters to produce various EOSs from this quarkyonic prescription. 

We vary $\Lambda$ and $\kappa$ over a range of $300-500$ MeV and $0.1-0.3$, respectively. The training set for the pSLM routine is composed of $18$ datasets, each composing $50$ snapshots consisting of radii, central pressures, masses and Love numbers. Using the pSLM algorithm, we predict the curves for test parameter sets as shown in Fig.~\ref{fig:parametricDMDs}, which were not used to train the emulator. 

The maximum mass and radius results from the HF solver and the pSLM emulator, as well as the percent error and speed of the calculation, are also recorded in Table~\ref{tab:eos_table}. The average speed for the pSLM emulator is $\approx 1.63 \times 10^{-4}$ seconds, yielding an overall average speed up of $\approx 7.0 \times 10^4$ over the HF RK4 solver. All percent errors are $< 1\%$ for all maximum masses and radii.


\section{Comparison to other solvers}
\label{sec:Compare}
To provide a comparison of the SLM and pSLM methods to other differential equation solvers we choose two finite difference methods, namely, the forward Euler method and RK2 (Heun's) method. These were chosen because they use one and two function evaluations per iteration, respectively, compared to the four used by RK4, and hence constitute a ``lower fidelity" solution. We note that all three of these solvers employ a fixed step size.
The TOV solutions using these methods are shown in Tables~\ref{tab:eos_table_euler}-~\ref{tab:eos_table_rk4}, where we compare them  to our SLM and pSLM results for each EOS.

To improve the accuracy of these low-fidelity (LF) solvers, twice the number of points were chosen. 
This results in an effective doubling of the runtime for the three solvers, as seen in Tables~\ref{tab:eos_table_euler}-\ref{tab:eos_table_rk4}. In Table~\ref{tab:eos_table_rk4}, we also show results for the HF RK4 method using double the number of points as in the previous sections to make a fair comparison between the methods in this subsection. As expected, there is a speedup of about three times in using the RK2 and Euler methods compared to the standard RK4 method. However, looking at the errors of these two methods compared to the SLM errors with respect to the HF solutions, SLM performs much better than the RK2 and Euler methods. Additionally, using these other, lower fidelity methods does not affect the order of magnitude of the SLM or pSLM runtimes.


\begin{table*}[htbp]
    \centering
    \begin{tabular}{|c|c|c|c|c|c|c|c|c|}
        \hline
        & \multicolumn{2}{c|}{Euler} & \multicolumn{2}{c|}{SLM} & \multicolumn{2}{c|}{Rel. Error (\%)} & \multicolumn{2}{c|}{Time (s)} \\
        \hline
         \multirow{2}{*}{EOS} & Max. & Radius & Max. & Radius & Max. & \multirow{2}{*}{Radius} & \multirow{2}{*}{Euler} & \multirow{2}{*}{SLM}  \\
         & Mass [$M_\odot$] & [km] & Mass [$M_\odot$] & [km] & Mass & & & \\
         \hline
         \multicolumn{9}{|c|}{Tabular EOS} \\
         \hline
         SLy4 & 2.067 & 9.979 & 2.065 & 9.977 & 0.06 & 0.02 & 
         5.748 & $4.830\times 10^{-4}$ \\
         APR & 2.192 & 9.932 & 2.193 & 9.931 & 0.01 & 0.01 & 
         5.542 & $4.807\times 10^{-4}$ \\
         FSU Garnet & 2.066 & 11.547 & 2.066 & 11.548 & 0.00 & 0.01 &
         5.551 & $5.269\times 10^{-4}$ \\
         BL  & 2.082 & 10.294 & 2.083 & 10.297 & 0.00 & 0.03 &
         5.540 & $4.621\times 10^{-4}$ \\
         DS-CMF-5 & 2.023 & 11.755 & 2.031 & 11.548 & 0.44 & 1.76 &
         5.548 & $3.998\times 10^{-4}$ \\
         \hline
         \multicolumn{9}{|c|}{Quarkyonic EOS} \\
         \hline
         $\Lambda$=300.00, $\kappa$=0.26 & 2.887 & 14.482 & 2.645 & 12.936 & 0.77 & 2.55 &
         5.559 & $1.898\times 10^{-4}$ \\
         $\Lambda$=352.63, $\kappa$=0.12 & 2.666 & 12.566 & 2.321 & 10.585 & 0.33 & 0.81 &
         5.521 & $1.662\times 10^{-4}$ \\
         $\Lambda$=373.68, $\kappa$=0.23 & 2.476 & 11.467 & 2.470 & 11.639 & 0.25 & 1.06 &
         5.489 & $1.469\times 10^{-4}$ \\
         $\Lambda$=436.84, $\kappa$=0.14 & 2.329 & 10.616 & 2.902 & 14.520 & 0.52 & 0.08 &
         5.599 & $1.631\times 10^{-4}$ \\
         $\Lambda$=500.00, $\kappa$=0.28 & 2.223 & 10.12 & 2.281 & 10.434 & 2.59 & 2.53 &
         5.473 & $2.360\times 10^{-4}$ \\
         \hline
    \end{tabular}
    \caption{
    Similar to Table~\ref{tab:eos_table}, except the forward Euler method is used. The number of points has been doubled to ensure stable and accurate solutions.
    }
    \label{tab:eos_table_euler}
\end{table*}


\begin{table*}[htbp]
    \centering
    \begin{tabular}{|c|c|c|c|c|c|c|c|c|}
        \hline
        & \multicolumn{2}{c|}{RK2} & \multicolumn{2}{c|}{SLM} & \multicolumn{2}{c|}{Rel. Error (\%)} & \multicolumn{2}{c|}{Time (s)} \\
        \hline
         \multirow{2}{*}{EOS} & Max. & Radius & Max. & Radius & Max. & \multirow{2}{*}{Radius} & \multirow{2}{*}{RK2} & \multirow{2}{*}{SLM}  \\
         & Mass [$M_\odot$] & [km] & Mass [$M_\odot$] & [km] & Mass & & & \\
         \hline
         \multicolumn{9}{|c|}{Tabular EOS} \\
         \hline
         SLy4 & 2.067 & 10.046 & 2.067 & 10.054 & 0.01 & 0.08 & 
         11.075 & $4.592\times 10^{-4}$ \\
         APR & 2.193 & 9.999 & 2.193 & 9.998 & 0.01 & 0.01 & 
         11.362 & $5.550\times 10^{-4}$ \\
         FSU Garnet & 2.066 & 11.607 & 2.067 & 11.607 & 0.03 & 0.00 &
         11.105 & $4.551\times 10^{-4}$ \\
         BL  & 2.083 & 10.361 & 2.083 & 10.361 & 0.01 & 0.00 &
         11.145 & $5.572\times 10^{-4}$ \\
         DS-CMF-5 & 2.023 & 11.815 & 2.031 & 11.611 & 0.41 & 1.73 &
         11.542 & $3.970\times 10^{-4}$ \\
         \hline
         \multicolumn{9}{|c|}{Quarkyonic EOS} \\
         \hline
         $\Lambda$=300.00, $\kappa$=0.26 & 2.887 & 14.536 & 2.668 & 13.011 & 0.09 & 3.14 &
         11.433 & $1.459\times 10^{-4}$ \\
         $\Lambda$=352.63, $\kappa$=0.12 & 2.666 & 12.626 & 2.321 & 10.649 & 0.35 & 0.21 &
         11.449 & $1.583\times 10^{-4}$ \\
         $\Lambda$=373.68, $\kappa$=0.23 & 2.476 & 11.527 & 2.450 & 11.668 & 1.07 & 1.31 &
         11.273 & $1.769\times 10^{-4}$ \\
         $\Lambda$=436.84, $\kappa$=0.14 & 2.329 & 10.683 & 2.891 & 14.585 & 0.14 & 0.36 &
         11.095 & $1.471\times 10^{-4}$ \\
         $\Lambda$=500.00, $\kappa$=0.28 & 2.224 & 10.187 & 2.281 & 10.495 & 2.60 & 3.13 &
         11.074 & $2.100\times 10^{-4}$ \\
         \hline
    \end{tabular}
    \caption{
    Similar to Table~\ref{tab:eos_table}, except the RK2 (Heun's) method is used. As in Table~\ref{tab:eos_table_euler}, the number of points has been doubled to ensure stable and accurate solutions.
    }
    \label{tab:eos_table_rk2}
\end{table*}


\begin{table*}[htbp!]
    \centering
    \begin{tabular}{|c|c|c|c|c|c|c|c|c|}
        \hline
        & \multicolumn{2}{c|}{RK4} & \multicolumn{2}{c|}{SLM} & \multicolumn{2}{c|}{Rel. Error (\%)} & \multicolumn{2}{c|}{Time (s)} \\
        \hline
         \multirow{2}{*}{EOS} & Max. & Radius & Max. & Radius & Max. & \multirow{2}{*}{Radius} & \multirow{2}{*}{RK4} & \multirow{2}{*}{SLM}  \\
         & Mass [$M_\odot$] & [km] & Mass [$M_\odot$] & [km] & Mass & & & \\
         \hline
         \multicolumn{9}{|c|}{Tabular EOS} \\
         \hline
         SLy4 & 2.067 & 10.033 & 2.066 & 10.034 & 0.03 & 0.02 & 
         22.519 & $4.580\times 10^{-4}$ \\
         APR & 2.193 & 9.986 & 2.195 & 9.987 & 0.10 & 0.01 & 
         22.515 & $4.930\times 10^{-4}$ \\
         FSU Garnet & 2.066 & 11.601 & 2.066 & 11.601 & 0.00 & 0.00 &
         22.629 & $4.508\times 10^{-4}$ \\
         BL  & 2.083 & 10.348 & 2.083 & 10.352 & 0.01 & 0.05 &
         22.536 & $4.742\times 10^{-4}$ \\
         DS-CMF-5 & 2.023 & 11.808 & 2.034 & 11.599 & 0.57 & 1.78 &
         22.507 & $4.930\times 10^{-4}$ \\
         \hline
         \multicolumn{9}{|c|}{Quarkyonic EOS} \\
         \hline
         $\Lambda$=300.00, $\kappa$=0.26 & 2.887 & 14.529 & 2.660 & 12.993 & 0.21 & 3.00 &
         23.135 & $1.621\times 10^{-4}$ \\
         $\Lambda$=352.63, $\kappa$=0.12 & 2.666 & 12.612 & 2.320 & 10.643 & 0.39 & 0.27 &
         22.808 & $1.569\times 10^{-4}$ \\
         $\Lambda$=373.68, $\kappa$=0.23 & 2.476 & 11.520 & 2.452 & 11.661 & 0.98 & 1.25 &
         22.607 & $1.438\times 10^{-4}$ \\
         $\Lambda$=436.84, $\kappa$=0.14 & 2.329 & 10.669 & 2.892 & 14.582 & 0.17 & 0.34 &
         22.751 & $1.359\times 10^{-4}$ \\
         $\Lambda$=500.00, $\kappa$=0.28 & 2.224 & 10.173 & 2.286 & 10.488 & 2.79 & 3.06 &
         22.797 & $2.222\times 10^{-4}$ \\
         \hline
    \end{tabular}
    \caption{
    Similar to Table~\ref{tab:eos_table}, except the HF RK4 method uses twice the number of points than those in Table~\ref{tab:eos_table} for a direct comparison with Tables~\ref{tab:eos_table_euler} and \ref{tab:eos_table_rk2}.
    }
    \label{tab:eos_table_rk4}
\end{table*}


\section{Summary and outlook \label{sec:summary}}

In this work, we presented SLM as a method to emulate the TOV equations, including tidal deformabilities. We employed SLM to emulate $M-R$ curves from tabular EOSs, such as the well-known SLy4 EOS. We then used a parametric version of SLM (pSLM) to emulate EOSs that possess an arbitrary number of parameters. In particular, we focused on the quarkyonic EOS. The results from these emulators agree well with HF RK4 evaluations of the TOV equations. We gain an average computational speed-up of $\sim 4.7 \times 10^4$ across both types of EOSs considered; hence, both SLM algorithms present highly efficient emulators for large-scale computations of the TOV equations in scenarios such as those presented by multi-messenger astrophysical inference frameworks. This emulation strategy will help to reduce the bottleneck that the TOV equations present in such calculations, without sacrificing necessary accuracy. 

Our SLM algorithm is not limited to small parameter sets, and as such can be easily generalized to EOSs with larger parameter sets. More broadly, the SLM algorithm works well for problems that possess logarithmic dependence in the solutions. Hence, our work can be straightforwardly extended in future to problems that share this trait, e.g., density functionals with non-affine parameter dependence, or time-dependent nuclear structure observables. To help other researchers utilize and build upon our work, the code that contains the SLM algorithm and that which produced the results in this paper, as well as tutorials to assist new users, is published in a public GitHub repository for the use of the scientific community~\cite{DMDGitHub}.


\begin{acknowledgments}
We thank J\'er\^ome Margueron for a careful review of our manuscript, Xilin Zhang for engaging discussions, and Amy L. Anderson for providing the FSUGarnet data. We also thank {\tt CompOSE} (\url{https://compose.obspm.fr}) for providing all other tabular EOS datasets. A.C.S. thanks the Facility for Rare Isotope Beams for their hospitality and encouragement during the completion of this work. This work is supported by the CSSI program, Award OAC-2004601 (BAND collaboration \cite{BAND_Framework}) (A.C.S.), and the U.S. Department of Energy, Office of Science, Nuclear Physics, under Award DE-SC0023688 (S.L.), Award DE-FG02-93-40756 (A.C.S.) and Award DE-SC0024233 (STREAMLINE collaboration) (J.M.M.).
\end{acknowledgments}


\appendix

\section{Scaling TOV and Tidal Equations}\label{TOVdetails}
The TOV equations~\cite{tolman1939, oppenheimer1939}, along with the tidal deformability equation~\cite{Hinderer:2007mb, Hinderer:2009ca, Postnikov:2010yn}, are given by\footnote{For a detailed derivation of these equations, see Refs.~\cite{Fantina:2022xas, Hinderer:2007mb, Hinderer:2009ca, Postnikov:2010yn}.}
\begin{align}
    \frac{dP}{dr} &= -\frac{G}{c^2}\left[\epsilon(r) + P(r)\right] \frac{m(r) + 4\pi r^3 P(r)/c^2}{r\left[r-2Gm(r)/c^2\right]}; \\
    \frac{dm}{dr} &= 4\pi r^2 \frac{\epsilon(r)}{c^2}; \\
    \frac{dy}{dr} &= -\frac{y(r)^2}{r} - \frac{F(r) y(r)}{r}  - \frac{Q(r)}{r},
\end{align}
where,
\begin{align}
    F(r) = \frac{1 - 4\pi G r^2 \left[\epsilon(r) - P(r) \right]/c^4}{1 - 2\frac{Gm(r)}{rc^2}}, 
\end{align}
and
\begin{equation}
\begin{aligned}
   Q(r) &= \frac{4\pi G r^2/c^4}{1- \frac{2Gm(r)}{rc^2}} \left( 5 \epsilon(r) + 9 p(r) + \frac{\epsilon(r) + p(r)}{c_s(r)^2}c^2 - \right. \\
   &\left. \frac{6c^4}{4\pi r^2 G} \right)
   - 4 \left(\frac{G\left[m(r)/(rc^2) + 4\pi r^2 p(r)/c^4\right]}{1 - 2 Gm(r)/(rc^2)} \right)^2.
\end{aligned}    
\end{equation}

Following closely Ref.~\cite{Piekarewicz2017}, we use the scaling $r = R_0x$, $m = M_0 m(x)$, $P = P_0 p(x)$, $\epsilon = \varepsilon_0 \varepsilon(x)$. To define the relevant scales, we set
\begin{equation}
    \varepsilon_0 = P_0 \equiv \frac{1}{8\pi^2} \frac{(m_nc^2)^4}{(\hbar c)^3} \approx 1.285\, \mathrm{GeV/fm^3}
\end{equation}
and adopt the natural normalization
\begin{equation}
    \left[\frac{2GM_0}{c^2R_0} \right] = \left[\frac{4\pi R_0^3\varepsilon_0}{3M_0c^2} \right] = 1.
\end{equation}
Setting these quantities to unity, meaning to a dimensionless value of order one, establishes \lq\lq natural" length and mass scales for the problem. This leads to
\begin{align}
    R_0 &= \sqrt{\frac{3\pi}{\alpha_G}} \lambda_n \approx 8.378\, \mathrm{ km}, \\
    M_0 &= \left(\frac{R_0}{R_s^{\odot}}\right) M_{\odot} \approx 2.837\, M_{\odot},
 \end{align}
where $\alpha_G$ is the small, dimensionless gravitational coupling strength between two neutrons (or the neutron mass to Planck mass ratio), \(\lambda_n\) is the Compton wavelength of the neutron, and \(R_s^{\odot}\) represents the Schwarzschild radius of the Sun. Numerically, these values are:
 \begin{align}
     \alpha_G &= \frac{G m_n^2}{\hbar c} \approx 5.922 \times 10^{-39} \\
     \lambda_n &= \frac{\hbar c}{m_n c^2} \approx 0.210 \times 10^{-18}\, \mathrm{km} \\
     R_s^{\odot} &= \frac{2GM_{\odot}}{c^2} \approx 2.953\, \mathrm{km}.
 \end{align}

The resulting scaled equations are then given by 
\begin{align}
    \frac{dp}{dx} & = -\frac{1}{2} \frac{\left[\varepsilon(x) + p(x)\right]\left[m(x) + 3x^3 p(x)\right]}{x^2\left[1 - m(x)/x\right]} \\
    \frac{dm}{dx} &= 3x^2 \varepsilon(x) \\
    \frac{dy}{dx} &= -\frac{y(x)^2}{x} - \frac{F(x)y(x)}{x} - \frac{Q(x)}{x} 
\end{align}
with
\begin{equation}
F(x) = \frac{1 - \tfrac{3}{2} x^{2} \left[\varepsilon(x) - p(x) \right]}{1-\frac{m(x)}{x}}, \nonumber
\end{equation}
and 
\begin{align*}
    Q(x) &= \frac{\tfrac{3}{2} x^2}{\left(1 - \frac{m(x)}{x} \right)} \left[5\varepsilon(x) + 9p(x) + \frac{\varepsilon(x) + p(x)}{c_s^2(x)} - \frac{4}{x^2} \right] \\
    &- \left(\frac{m(x)/x + 3 x^2 p(x)}{\left(1 - \frac{m(x)}{x} \right)} \right)^2.
\end{align*}

Solving these equations, we get the scaled $m(x),\, p(x),\, y(x)$. To calculate the mass-radius curve of a neutron star, one enforces the boundary condition that the pressure of star is zero at the surface. This enables one to find the mass and radius of the star for a given central pressure.

We use $y_R = y(r=R)$ where $R$ is the radius of the star with mass $M [M_{\odot}]$ to calculate the compactness of the star $\beta = G M/R$. Using this one can then calculate the dimensionless Love number $k_2$ (see Eq.~\eqref{eq:lovenumber}).

\section{Banach greedy recombination interpolation method (B-GRIM) }\label{app:BGRIM}
In this approach~\cite{lyons2024}, the goal is to find sparse approximations of functions thereby reducing computational complexity. The objective is to construct an interpolant that minimizes the norm ($\ell^1$) in that space.  In the greedy recombination method, the algorithm iteratively selects basis elements to improve the interpolation, minimizing the error at each step. The method is composed of  two main steps: 1) Banach extension and 2) Banach recombination.

\subsection{Banach Extension}
Let $u \in \mathrm{Span}(\mathcal{F})$ where $\mathcal{F}$ is the set of functionals in the training set corresponding to features that approximate function $\bm{y}$, and assume $L \subset \Sigma$ be a finite collection of linear functionals, where $\Sigma$ denotes the data. For each subset $L \subset \Sigma$, the method employs a recombination process to find an approximation $u \in \mathrm{Span}(\mathcal{F})$ for the target function $\bm{y}$, where $\bm{y}$ represents the desired solution constrained by $L$. Typically this process introduces numerical errors; to minimize these, it is assumed that $u$ is sufficiently close to $\bm{y}$ at each functional $\sigma \in L$, a relationship that is formalized in the recombination step. 

\subsection{Banach recombination}
Banach recombination phase consists of three steps. 
\begin{enumerate}
    \item Choose a subset $L_j \subset \Sigma$ by randomly permuting the element order in $L$. 
    \item Perform recombination thinning to find an element $u_j \in \mathrm{Span}(\mathcal{F})$ that satisfies $|\sigma (\bm{y} - u_j) | \leq \varepsilon_0$ for every $\sigma \in L_j$, where $\varepsilon_0$ is a small tolerance value.
    \item Compute the error metric 
    $E[u_j] := \max\{|\sigma(\bm{y} -u_j)|: \sigma \in \Sigma \}$.
\end{enumerate}
After identifying the elements $u_1,\ldots, u_s$, define $u$ as
\begin{equation}
    u := \mathrm{argmin}\,{E[w] : w \in \{u_1, \ldots, u_s \}}
\end{equation}
This final $u$ serves as the best approximation of the target function $\bm{y}$. The full algorithm and derivation of the method is provided in Ref.~\cite{lyons2024}.

\bibliography{refs}

\end{document}